\documentclass[%
 reprint,
superscriptaddress,
 amsmath,amssymb,
 aps,
]{revtex4-1}

\usepackage{graphicx}
\usepackage{dcolumn}
\usepackage{bm}
\usepackage{siunitx}
\usepackage{hyperref}

\begin{document}

\title{Nonlocal signatures of hybridization between\\ quantum dot and Andreev bound states}%

\author{Andreas P\"oschl}
\affiliation{Center for Quantum Devices, Niels Bohr Institute, University of Copenhagen, 2100 Copenhagen, Denmark}

\author{Alisa Danilenko}
\affiliation{Center for Quantum Devices, Niels Bohr Institute, University of Copenhagen, 2100 Copenhagen, Denmark}

\author{Deividas Sabonis}
\affiliation{Center for Quantum Devices, Niels Bohr Institute, University of Copenhagen, 2100 Copenhagen, Denmark}
\affiliation{Laboratory for Solid State Physics, ETH Z\"urich, CH-8093 Z\"urich, Switzerland}%

\author{Kaur Kristjuhan}
\affiliation{Center for Quantum Devices, Niels Bohr Institute, University of Copenhagen, 2100 Copenhagen, Denmark}%

\author{Tyler Lindemann}
\affiliation{Department of Physics and Astronomy, and Birck Nanotechnology Center, Purdue University, West Lafayette, Indiana 47907 USA}

\author{Candice Thomas}
\affiliation{Department of Physics and Astronomy, and Birck Nanotechnology Center, Purdue University, West Lafayette, Indiana 47907 USA}

\author{Michael J. Manfra}
\affiliation{Department of Physics and Astronomy, and Birck Nanotechnology Center, Purdue University, West Lafayette, Indiana 47907 USA}
\affiliation{School of Materials Engineering, and School of Electrical and Computer Engineering, Purdue University, West Lafayette, Indiana 47907 USA}

\author{Charles M. Marcus}
\affiliation{Center for Quantum Devices, Niels Bohr Institute, University of Copenhagen, 2100 Copenhagen, Denmark}

\begin{abstract}
We investigate local and nonlocal signatures of hybridization between a quantum dot state and an extended Andreev bound state (ABS) in a gate-defined InAs nanowire with multiple side probes. When a quantum dot in one of the side probes was hybridized with an ABS in the nanowire, a characteristic spectroscopic pattern was observed both locally, i.e., in the probe with the quantum dot, and nonlocally, in the tunnel conductance of a remote probe. Nonlocal signatures of hybridization reveal the extended nature of the ABS.
\end{abstract}

\maketitle

Progress in material growth has enabled the realization of hybrid materials with distinct low-temperature phases \cite{krogstrup_epitaxially,Yu_AlEuSInAs,krizek_SAG,Pereira2021} not observed in the constituent bulk materials \cite{FUKANE, FUKANE2, CTS_from_proxyQHS, proxyQHS_islands_fiboanyon, oreg_tripletCNT, magneticinsul_SDS, cook_proxyTIwires}. An important example is a superconductor grown epitaxially on a semiconductor having gateable carrier density, large negative $g$-factor, and strong spin-orbit coupling \cite{Lutchyn2018, Carrad2020, Yuan2021}. A promising material platforms that allows for scalable fabrication of advanced devices in this context are InAs two-dimensional electron gases (2DEGs) proximitized by superconducting Al \cite{javad_2deg,henri_lead,morten_sqpcn,alex_QDinLoop,eoins_dots,nichele_scaling,alex_interfero}. Devices of suitable geometries allow exploration of various bound states in nanowires (NWs), including Yu-Shiba-Rusinov states, Andreev bound states (ABSs), and Majorana bound states \cite{alex_QDinLoop, chang_YSR, kasper_tuningYSR,lee_scaling,pillet_CNT,nichele_ABS_Ic,nichele_scaling,dassama_meta}. 

The use of semiconductor-superconductor hybrids facilitates the realization of electrostatically controlled quantum dots (QDs) coupled to superconductors. QDs coupled to ABSs have received considerable attention from theoretical studies, including the use of the QD as a tool for measuring bound state lifetimes \cite{leijnse_dotMBSlifetime} or providing Majorana parity readout \cite{gharavi_dotMBSparity,hoffman_dotMBScomputation,karsten_dotMBS_operations,leijnse_QItransferDotMBS,plugge_boxqubits, alex_interfero}. In particular, the hybridization between a QD and a bound state leads to a shift of the bound state energy resulting in a characteristic `bowtie' or `diamond' shape. This can be used to determine the nonlocality and the spin structure of bound states \cite{clarke_quality, elsa_nonlocality, mingtang_nonlocality, elsa_quantifying}.  

In this Letter, we perform tunneling spectroscopy of a NW in a novel geometry that allows measurements at several side branches along the NW length using electrostatic gates patterned on an InAs/Al hybrid heterostructure. A similar configuration has been investigated theoretically \cite{dassarma_circuits}, and a related experiment has been carried out in a conventional nanowire with deposited superconductor and normal metallic side contacts \cite{grivnin_multiprobe}. In addition to ABSs due to bound states in the NW, we find conductance resonances due to accidental QDs in the tunnel barriers. We investigate hybridization of QD states with ABSs in the NW, observing signatures of hybridization both locally, that is, at the position of the accidental QD, and nonlocally, measured on another side probe away from the QD.

\begin{figure*}[t]
\includegraphics[scale=0.75]{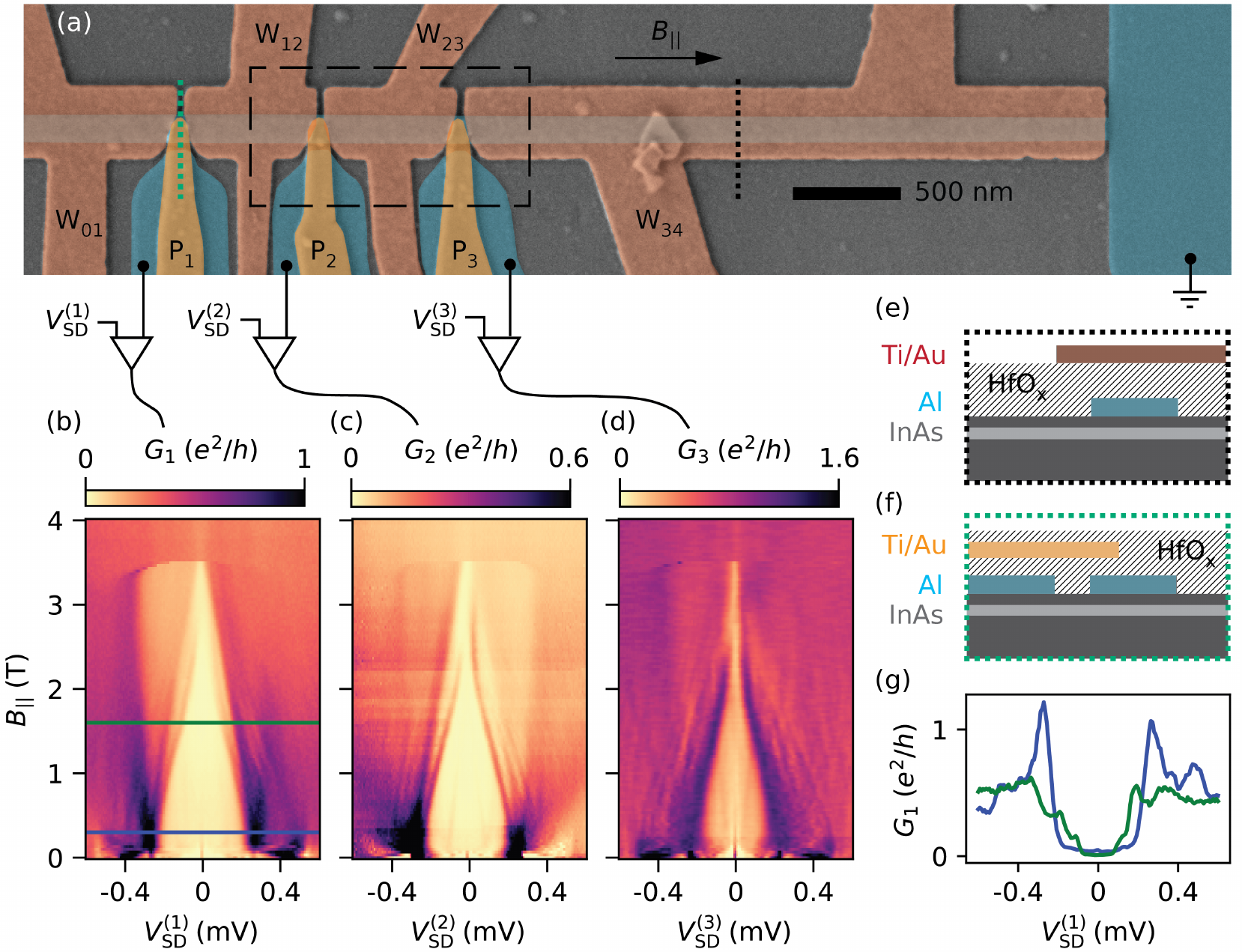}
\caption{\label{fig:device}(a) False colored scanning electron micrograph of device 1. The device consists of patterned epitaxial Al, forming a long narrow nanowire with several tunnel probes on top of an InAs quantum well. Gates labeled $\mathrm{P}_j$ are used to tune the tunnel barrier between the tunnel probe $j$ and the wire ($j \in \{1,2,3\}$). Gates labeled $\mathrm{W}_{kl}$ with $kl \in \{01,12,23,34\}$ deplete carriers except under the Al. (b-d) Tunneling spectroscopy at three probes with all gate voltages $V_{\mathrm{W}kl}=\SI{-4.5}{\volt}$. (e, f) Schematic cross sections of the device at the positions given by the green and black dotted lines in (a). (g) Line cuts at field values $B_{||}=\SI{0.3}{\tesla}$ and $B_{||}=\SI{1.6}{\tesla}$ indicated by the blue and green line in (b).}
\end{figure*}

Figure \ref{fig:device}(a) shows a micrograph of device 1, based on an InAs 2DEG with \SI{5}{\nano\meter} of epitaxial Al. The device consists of an Al strip of width \SI{100}{\nano\meter} and length \SI{5}{\micro\meter}, connected at both ends to large planes of Al that were electrically grounded. Gates labeled $\mathrm{W}_{kl}$ were Ti/Au on top of \SI{30}{\nano\meter} $\mathrm{HfO_2}$, as shown in Fig.~\ref{fig:device}(e). Gates were used to deplete the semiconductor on either side of the Al wire, creating by depletion a quasi one-dimensional InAs NW self-aligned to the proximitizing Al.

Neighboring gates $\mathrm{W}_{kl}$ form a constriction that acts as a tunnel probe. The lead of the probe, away from the tunneling region, is made using the same unetched epitaxial Al. Tunneling across the bare semiconductor region between Al NW the Al lead is controlled by a probe gate,  $\mathrm{P}_j$, as shown in Fig.~\ref{fig:device}(f). Details of the materials and fabrication are given in the Supplementary Material. 

 The measurement set-up is shown schematically in Fig.~\ref{fig:device}(a). With the NW  grounded, individual voltage biases $V_{\mathrm{SD}}^{(j)}$ were applied on probe $j$ via current to voltage converters.
 Tunneling currents $I_j$ through the tunnel barriers were measured using lock-in detection yielding differential conductances $G_j=\mathrm{d}I_j/\mathrm{d}V^{(j)}_{\mathrm{SD}}$. Measurements were carried out in a cryo-free dilution refrigerator with a 6-1-1 \SI{}{\tesla} vector magnet at $\approx\SI{15}{\milli\kelvin}$ mixing-chamber temperature. 
 
Tunneling conductances $G_j$ as a function of magnetic field $B_{||}$ applied parallel to the NW are shown in Figs.~\ref{fig:device}(b-d). For weak tunneling and in the absence of probe resonances, $G_j$ is proportional to the density of states in the NW. The superconducting gap of the Al in the leads of the probes closes at low field, $B_{||} \approx \SI{0.2}{\tesla}$ above which the probes can be regarded as normal metal, as discussed previously \cite{henri_lead, nichele_scaling}. The semiconductor under the Al in the NW was depleted by setting all $\mathrm{W}_{kl}$ gates to $\SI{-4.5}{\volt}$.  Measurements  on all three probes showed a superconducting gap closing without any subgap states crossing zero energy. For $G_1$ this is illustrated by the line cuts in Fig.~\ref{fig:device}(g). Note that the measurement of $G_3$ shows finite subgap conductance, which we attribute to probe 3 being tuned to an open regime with high-bias conductance $G_3(V^{(3)}_{\mathrm{SD}}=\SI{0.4}{\milli\volt}) \gtrsim 1 \; e^2/h$.\\

\begin{figure}
\includegraphics[scale=0.9]{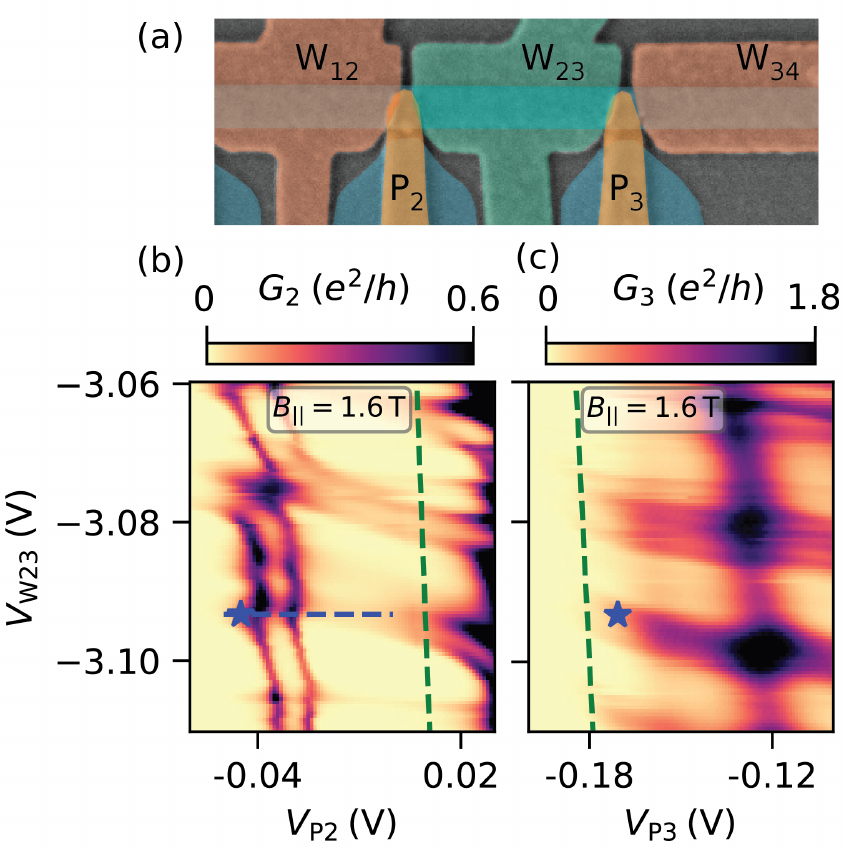}
\caption{\label{fig:gatemap}(a) Micrograph of the NW segment under investigation. $V_\mathrm{W12}=V_\mathrm{W34}=\SI{-7.0}{\volt}$ while the voltage on gate $\mathrm{W}_{23}$ (green), is varied. (b), (c) Differential conductance at zero bias measured at the left and the right end of the NW segment. Horizontal conductance resonances appear in both maps at similar gate voltages. Vertical conductance features, strongly dependent on gates $V_\mathrm{P2}$ and $V_\mathrm{P3}$ which tune the tunnel barriers, are also visible.
}
\end{figure}

To investigate the hybridization of a probe QD state with an ABS in the NW, we focus on the \SI{0.6}{\micro\meter} long NW segment under gate $\mathrm{W_{23}}$, see dashed box in Fig.~\ref{fig:device}(a), shown in Fig.~\ref{fig:gatemap}(a). To create an ABS in this segment, the voltage on gate $\mathrm{W_{23}}$ was set less negative, in the range of $\SI{-3}{\volt}$, while voltages on neighboring gates $\mathrm{W_{12}}$ and $\mathrm{W_{34}}$ were set to $\SI{-7.0}{\volt}$. At $B_{||}=\SI{1.6}{\tesla}$ and zero source drain biases, $V_{\mathrm{SD}}^{(j)}=0$, conductances $G_2$ and $G_3$ were measured as functions of  probe-gate voltages $V_\mathrm{P2}$ and $V_\mathrm{P3}$, respectively, and wire-gate voltage $V_\mathrm{W23}$. For both tunnel junctions, two sets of conductance resonances can be distinguished in Figs.~\ref{fig:gatemap}(b) and (c) by their characteristic slope. 
The first set primarily consists of vertical features that are strongly dependent on the gate voltages $V_\mathrm{P2}$ ($V_\mathrm{P3}$), which we attribute to  QDs in the tunnel barriers. The second set are predominantly horizontal, depending more strongly on $V_\mathrm{W23}$. The latter resonances are visible in both $G_2$ and $G_3$, suggesting that they arise from ABSs that extend over the segment covered by gate $\mathrm{W}_{23}$. Figure S1 in the Supplementary Material shows the complete evolution of tunneling spectroscopy from $V_\mathrm{W23} = \SI{-3.8}{\volt}$ to $V_\mathrm{W23} = \SI{-3.0}{\volt}$ while keeping the tunnel barriers at roughly constant transparency. This was achieved by compensating the effect of the gate $\mathrm{W}_{23}$ on the tunnel barriers by 
changing the gate voltages $V_\mathrm{P2}$, $V_\mathrm{P3}$ given by the green dashed line in Figs.~\ref{fig:gatemap}(b, c). 

\begin{figure}
\includegraphics[scale=0.9]{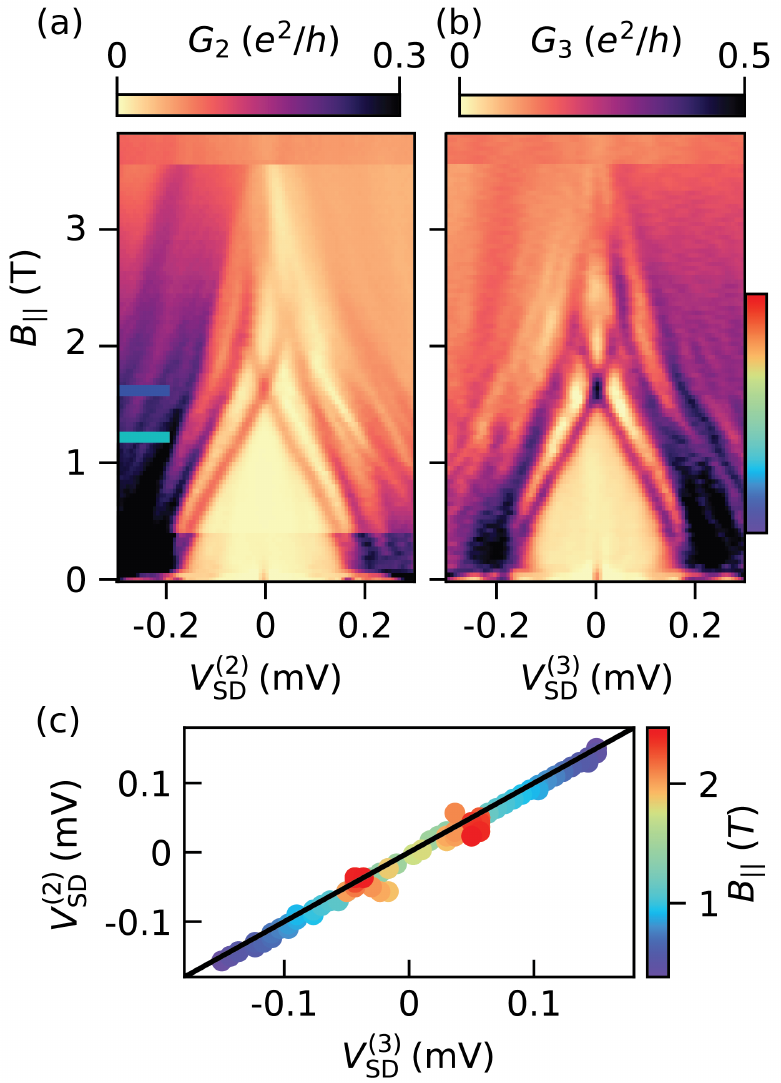}
\caption{\label{fig:fieldscan}(a, b) Tunneling spectroscopy with respect to magnetic field at the two ends of the NW with $V_\mathrm{W23}=\SI{-3.09}{\volt}$ [marked with $\star$ on Fig.~\ref{fig:gatemap}(b, c)]. Both measurements show subgap states crossing zero bias at $B_{||}=\SI{1.6}{\tesla}$ with a clear overshoot around $B_{||}=\SI{2}{\tesla}$. (c) Parametric plot of the extracted peak positions from the lowest energy subgap states in (a) and (b). 
The color of the points indicates the field value in accordance with the rainbow color bar in (b). }
\end{figure}

The blue star markers in Figs.~\ref{fig:gatemap}(b, c) at gate voltages $V_\mathrm{W23}=\SI{-3.09}{\volt}$, $V_\mathrm{P2}=\SI{-0.045}{\volt}$, and $V_\mathrm{P3}=\SI{-0.170}{\volt}$ mark ABSs that are weakly tunnel coupled to the probes. Tunneling spectroscopy of these ABSs as a function of magnetic field $B_{||}$ in Figs.~\ref{fig:fieldscan}(a, b) reveals a zero-bias crossing of the ABSs at $B_{||}=\SI{1.6}{\tesla}$ followed by an overshoot at $B_{||}=\SI{2}{\tesla}$. The states appear in both tunneling conductance measurements of $G_2$ and $G_3$. We extracted the peak position in $V^{2(3)}_{\mathrm{SD}}$ of the ABS from the measurements of $G_2$ and $G_3$. 
The parametric plot of the peak positions $V^{(2/3)}_{\mathrm{SD}}$ of the ABSs in $G_2$ and $G_3$ in Fig.~\ref{fig:fieldscan}(c) shows that all points lie close to the identity line, suggesting strong correlations. Details about the extraction of the peak position are outlined in the Supplementary Material.

The ABSs seen in $G_2$ and $G_3$ evolve similarly with gate voltage $V_\mathrm{W23}$ and magnetic field $B_{||}$, suggesting that they belong to the same extended quantum states. Similar experimental findings have been made previously \cite{gian_correlations,frolov_quantized}. The magnetic field dependence of the states is furthermore characteristic for ABSs in short NWs \cite{dassama_meta}.\\

\begin{figure}
\includegraphics[scale=0.9]{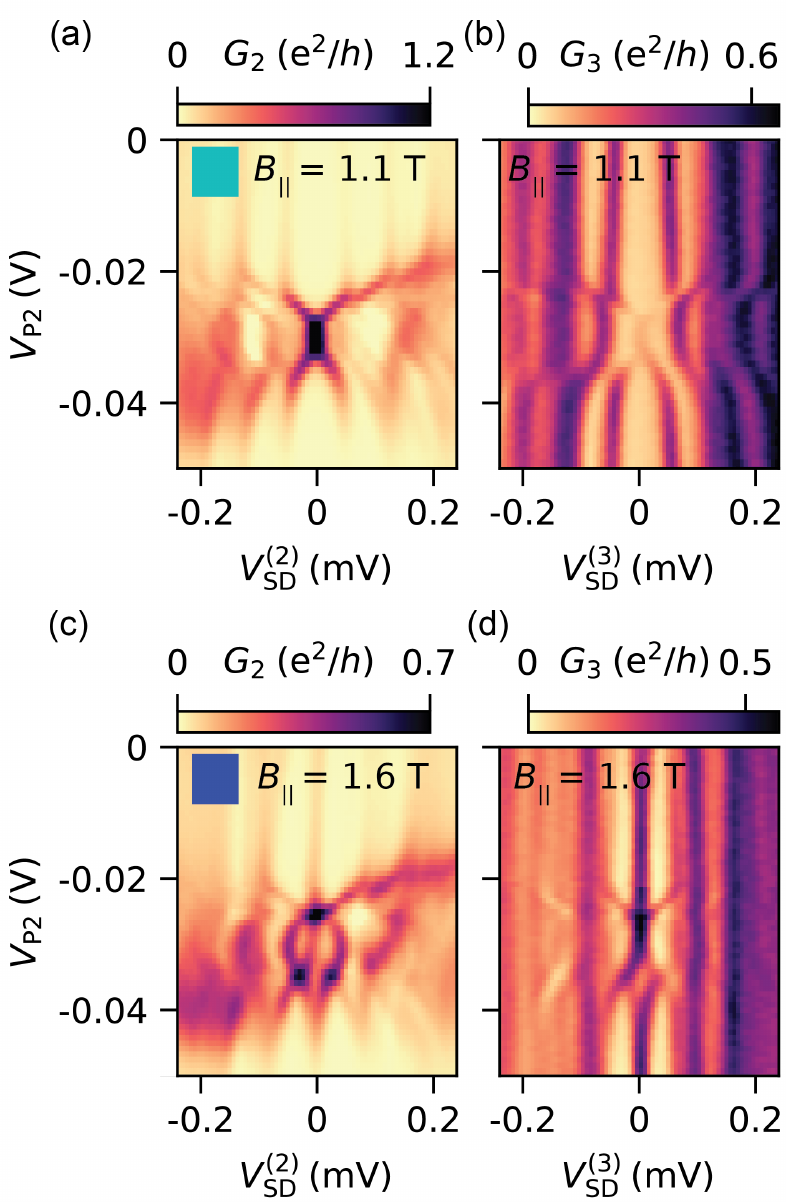}
\caption{\label{fig:pc}(a) Tunneling conductance $G_2$ at the left side of the NW segment at $V_\mathrm{W23}=\SI{-3.09}{\volt}$ [marked with $\star$ and blue dashed line on Fig.~\ref{fig:gatemap}(b, c)] as a function of the gate voltage $V_\mathrm{P2}$ which tunes the tunnel barrier. 
A QD resonance is visible as an enhancement of conductance at high bias around $V_2\approx\SI{-0.030}{\volt}$. The subgap states change their energy at the point of the QD resonance, drawing a characteristic `bowtie' shape. 
(b) Tunneling spectroscopy $G_3$ at the other end of the NW. The ABS show the same change in energy as visible in the measurement of $G_2$ in (a). (c, d) same as (a, b) at higher parallel magnetic field, $B_{||}$. The ABSs split to form a `diamond' shaped energy profile at the position of the QD resonance.}
\end{figure}

Special points in the measurement in Fig.~\ref{fig:gatemap}(b) are the crossing points of the horizontal resonances with the sharp vertical resonances. At these points, an ABS in the NW is on resonance with the QD in the tunnel barrier under the gate $\mathrm{P_2}$. Tunnelling spectroscopy $G_2$ using tunnel probe 2 at a field value of $B_{||}=\SI{1.1}{\tesla}$ while sweeping $V_\mathrm{P2}$ along the values given by the blue dashed line in Fig.~\ref{fig:gatemap}(b) is shown in Fig.~\ref{fig:pc}(a). The ABSs were unaffected by the change of $V_\mathrm{P2}$ outside the range \SIrange{-0.040}{-0.020}{\volt}. Within this range, the QD resonance appears as a conductance enhancement at high bias, reflecting the fact that $G_2$ was being measured through the QD in tunnel barrier 2. As the QD went on resonance with the ABSs, the ABSs with lowest energy merged at zero bias before returning to their previous energies. This resulted in a characteristic `bowtie' shape of the resonances of the ABSs. A simultaneous measurement of $G_3$ during the sweep of $V_\mathrm{P2}$ at the other end of the NW is shown in Fig.~\ref{fig:pc}(b). The enhancement of conductance at high bias due to the QD that was present in the measurement of $G_2$ was absent in the measurement of $G_3$. The ABSs, however, showed the same `bowtie' shape around the voltage value $V_\mathrm{P2}\approx\SI{-0.030}{\volt}$ which corresponds to the resonance condition between the ABSs and the QD in tunnel barrier 2. Note that in the measurement of both $G_2$ and $G_3$ not only the lowest energy ABSs undergoes a change at the resonance condition with the QD, but also the higher excited states. In addition to the change in ABS energy, a clear change in the conductance peak height is visible when going through the resonance condition.

Around $B_{||}=\SI{1.6}{\tesla}$, the ABSs merged to yield a single conductance peak at zero bias. A measurement of $G_2$ with respect to $V_\mathrm{P2}$ in Fig.~\ref{fig:pc}(c) shows that this peak was unperturbed except for voltage values around $V_\mathrm{P2}\approx\SI{-0.030}{\volt}$ where the QD was on resonance with the ABSs. At this point, the ABS resonances split symmetrically away from zero bias, forming a `diamond' shape. The simultaneous measurement of $G_3$ in Fig.~\ref{fig:pc}(d) reveals similar $V_\mathrm{P2}$ dependence of the ABS energy. Note that higher excited states were also affected around $V_\mathrm{P2}\approx\SI{-0.030}{\volt}$. \\

The appearance of ABSs with `bowtie'- and `diamond'-shaped patterns while on resonance with the QD level is an indication of the QD being sufficiently tunnel coupled to the ABSs such that the two energy levels significantly hybridize, consistent with theoretical and previous experimental results \cite{elsa_nonlocality, clarke_quality, vuik_quasiMBS, mingtang_nonlocality,elsa_quantifying}. The measurement of the energy shift at both ends of the \SI{0.6}{\micro\meter} NW, while a local gate voltage at only one end is changed, is a nonlococal signature of the delocalized ABSs. If tunneling spectroscopy is measured at both ends of a NW, hybridization of bound states with a local QD can be used as a quantum mechanical tool to test whether a quantum state extends through the whole NW, similar to the analysis of cross-conductance and correlated appearance at both ends \cite{gerbold_nonlocal, gian_correlations}. This is in contrast to experiments where spectroscopy is performed at one end of a NW. In such a case, a QD in the absence of a bound states in the NW can mimic signatures of extended states inside the NW in tunneling spectroscopy \cite{dassarma_against_mingtang, lee_singletdoublet, trivial_fullshell,dassarma_ABSvsMBSspectroscopy, dassarma_goodbadugly, vuik_quasiMBS}.

In comparison to previous experiments, the present set-up offers additional information about the spatial extent of the bound state, as one can perform tunneling spectroscopy at at both ends of the NW segment. This also allows for the observation of the change in energy of the ABS at one position while it is being hybridized with a QD \SI{0.6}{\micro\meter} away by the means of changing a local gate. 
This nonlocal signature is a demonstration of the ABS being an extended quantum state. 

In the Supplementary Material we present data of the hybridization of an ABS in the NW segment under the gate labeled $\mathrm{W}_{12}$ with a local QD in the tunnel barrier. Results from a second device (device 2) are also presented. Device 2 is a slightly different design, with side probes that do not have Al leads, only the bare semiconductor. 
Device 2 showed similar results as device 1 when an ABS in a NW was brought onto resonance with a QD localized at one probe while observing the impact of the hybridization on the bound state at the other probe.

\begin{acknowledgments}
We thank Samuel Escribano, Karsten Flensberg, Max Geier, Andrea Maiani, Elsa Prada, Pablo San-Jose, and Waldemar Svejstrup  for valuable discussions on theory, and Abhishek Banerjee, Lucas Casparis, Asbj\o rn Drachmann,  Esteban Martinez, Felix Passmann, Daniel Sanchez, Saulius Vaitiek\.enas, and Alexander Whiticar for input on experimental aspects. We acknowledge support from the Danish National Research Foundation, Microsoft, and a research grant (Project 43951) from VILLUM FONDEN.
\end{acknowledgments}

\bibliography{references_pc}

\appendix
\clearpage
\onecolumngrid

\begin{center}
{\bf SUPPLEMENTARY MATERIAL}
\end{center}

\setcounter{equation}{0}
\setcounter{figure}{0}
\setcounter{table}{0}
\setcounter{page}{1}
\makeatletter
\renewcommand{\theequation}{S\arabic{equation}}
\renewcommand{\thefigure}{S\arabic{figure}}

\section{Material, Fabrication, and Measurement Details}
\subsection{\label{app:wafer}Wafer information}
The material used for device 1 was an InGaAs/InAs/InAlAs heterostructure covered with an in-situ epitaxially grown Al top layer. Electron mobility in the InAs quantum well after removal of the Al by wet etching was measured using a Hall bar to be $\SI{25e3}{\centi\meter\squared\per\volt\per\second}$. The material for device 2 was an InAlAs/InAs/InAlAs heterostructure with a similar epitaxial Al top layer.

\subsection{\label{app:fab}Device Fabrication}
Device fabrication was performed as follows:  leads and bond pads were formed by wet etching a mesa structure defined by electron beam lithography (EBL). The wet etch was performed in two stages. First, the Al film was removed in the EBL patterned area using Transene aluminum etchant type D. Then, the semiconductor was etched $\sim \SI{350}{\nano\meter}$ deep using a solution of $\mathrm{H_2O:C_6H_8O7:H_3PO_4:H_2O_2}$ (220:55:3:3). Fine features (wire and probe leads) were patterned using a second EBL step. The Al film was again selectively wet etched using Transene aluminum etchant type D at $50^\circ\mathrm{C}$. For device 1 a first layer of gate dielectric made of \SI{15}{\nano\meter} $\mathrm{HfO_x}$ was grown globally on the chip using atomic layer deposition. The 5/20/5 \SI{}{\nano\meter} Ti/Au/Ti gates $\mathrm{P}_j$ were patterned by a third EBL step, followed by a metal deposition liftoff. Bond pads and connecting lines were fabricated in the same way with 10/350 \SI{}{\nano\meter} of Ti/Au. After a second global deposition of \SI{15}{\nano\meter} $\mathrm{HfO_x}$ gate dielectric, the $\mathrm{W}_{kl}$ gates made from 5/20 \SI{}{\nano\meter} Ti/Au were deposited in a fifth EBL step. The corresponding bond pads, and connecting lines were fabricated from 10/350 \SI{}{\nano\meter} Ti/Au in a sixth EBL step, again using metal evaporation and liftoff. Note that $\mathrm{P}_{j}$ gates are buried inside dielectric. This way, narrow gates $\mathrm{P}_{j}$ can be fabricated that are close to the $\mathrm{W}_{kl}$ gates without risking a galvanic connections. \\

For device 2 only a single layer of \SI{15}{\nano\meter} thick $\mathrm{HfO_x}$ was deposited using atomic layer deposition. All gates made from 5/20 \SI{}{\nano\meter} thick Ti/Au were fabricated using EBL, metal deposition, and liftoff. The bond pads and connecting lines to the gates were fabricated the same way from 10/350 \SI{}{\nano\meter} thick Ti/Au.\\

\subsection{\label{sec:el_setup}Measurement setup}

\begin{figure}
\includegraphics[scale=0.9]{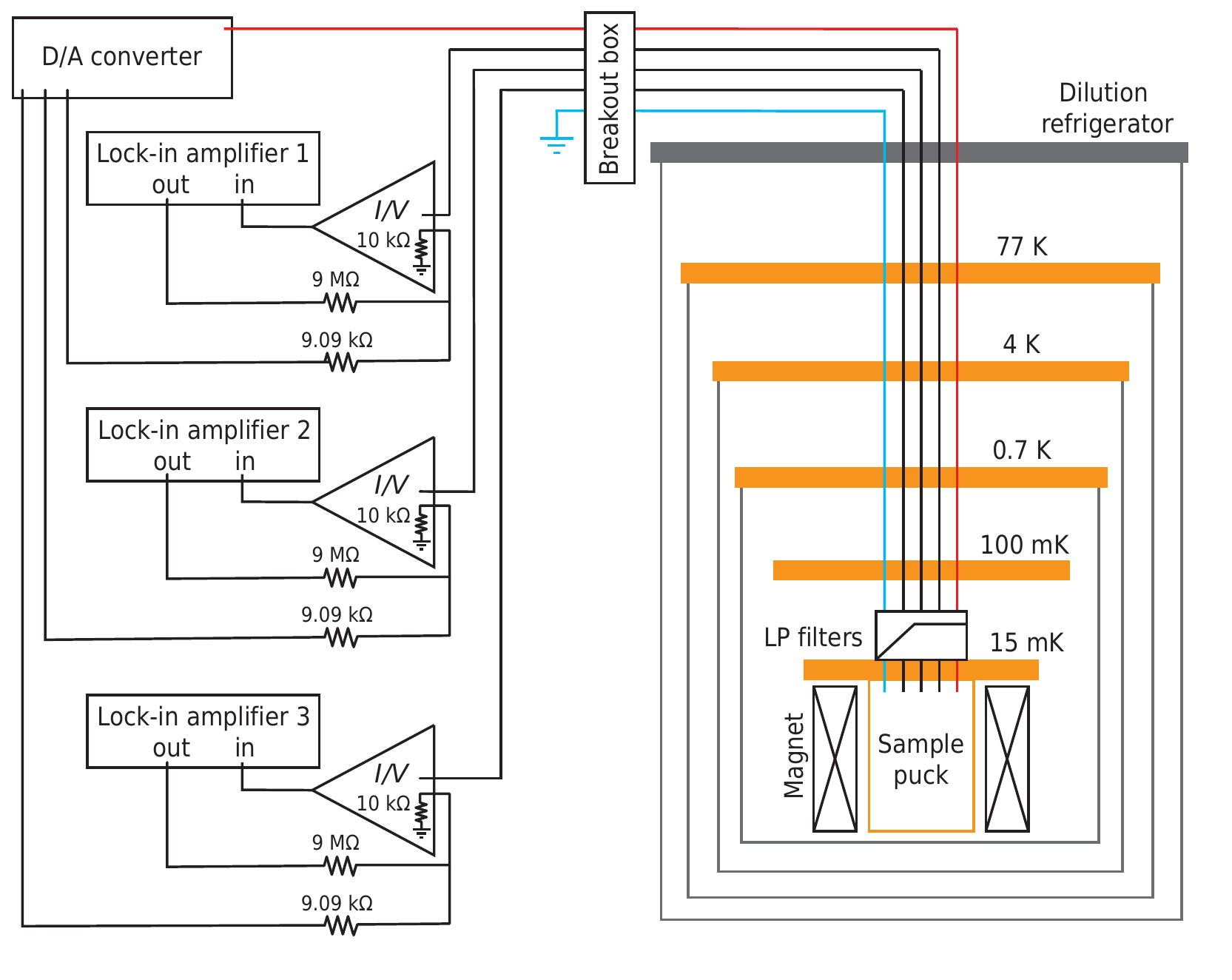}
\caption{\label{fig:setup} Schematic overview of the experimental measurement setup. All lines that supply gate voltages are pictured by the red line. The four lines that are used to connect the ground planes of the sample to the breakout box are represented by a single blue line. The dilution refrigerator is an Oxford Instruments Triton 400. The three current-to-voltage ($I/V$) converting amplifiers are made by Basel Precision Instruments (part number SP983c). Source-drain bias voltage is supplied to the sample via the input voltage offset port of the $I/V$ converters. Suitable resistors together with the input resistance of this port form a voltage divider. The lock-in amplifiers are made by Stanford Research Systems (model SR830 and SR860).}
\end{figure}

A schematic of the experimental set-up for transport measurements is depicted in Fig. \ref{fig:setup}. The sample was mounted in a puck-loading cryo-free dilution refrigerator \footnote{Oxford Instruments, Triton 400} equipped with a 6-1-1 \SI{}{\tesla} vector magnet. Throughout the measurements, the mixing chamber of the dilution fridge was at base temperature of roughly \SI{15}{\milli\kelvin}, without active control, as measured by a $\mathrm{RuO_2}$-based thermometer. Gate voltages were generated using an in-house built digital-to-analog converter with 20-bit precision. The three tunneling currents $I_j$ were amplified using a current-to-voltage converter \footnote{Basel Precision Instruments SP983c} followed by an AC lock-in amplifier \footnote{Stanford Research Systems SR830, SR860}. Bias voltages $V^{(j)}_{\mathrm{SD}}$ were applied via the offset voltage inputs of the current-to-voltage converters. The DC component of $V^{(j)}_{\mathrm{SD}}$ was supplied by the digital-to-analog (D/A) converter and an additional AC component $\mathrm{d}V^{(j)}_{\mathrm{SD}}$ at a frequency $f_j$ was supplied by the sine output of lock-in amplifier $j$. The frequencies were $f_1=\SI{42.2}{\hertz}$, $f_2=\SI{31.9}{\hertz}$, $f_3=\SI{20.4}{\hertz}$. In-house built multi-stage low-pass filters at the mixing chamber plate were used to attenuate electrical noise. The two ground planes of Al that connect to the nanowire (NW) at both ends were connected via two lines each to ground at the breakout box. 

For measurements in Figs.~1(b-d) in the main text, the DC bias voltage was swept simultaneously for all three tunnel probes, i.e., $V_{\mathrm{SD}}^{(1)}=V_{\mathrm{SD}}^{(2)}=V_{\mathrm{SD}}^{(3)}$. For all other tunneling spectroscopy measurements, where only two conductances are reported, the third probe was left floating. The two bias voltages were swept sequentially, with the respective other set to zero.

\section{\label{sec:S1}Additional data, device 1}
\subsection{\label{sec:plunger_scans}$\mathbf{W_{23}}$ dependence of subgap states}
\begin{figure}
\includegraphics[scale=1.0]{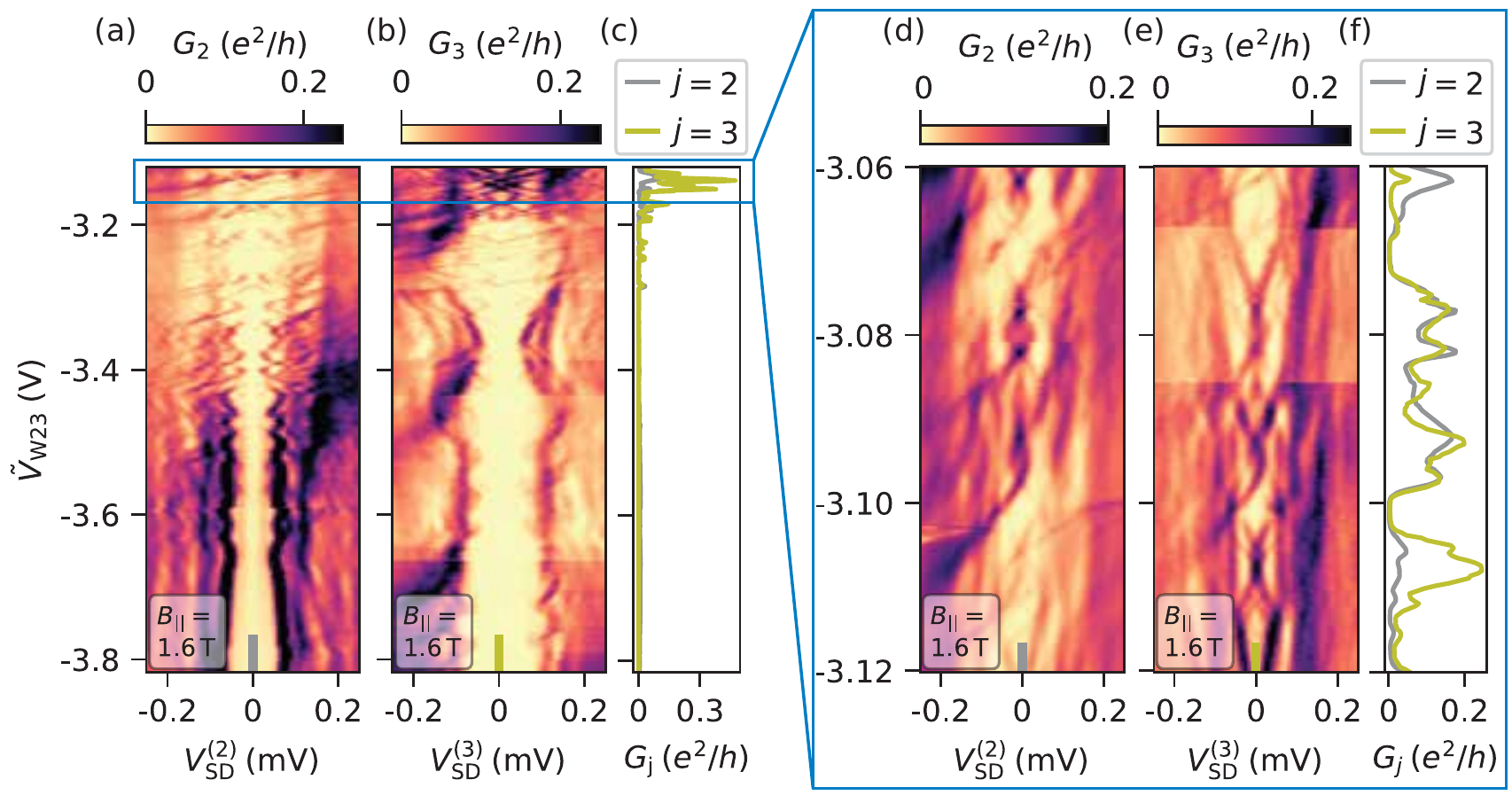}
\caption{\label{fig:evolution}(a, b) show conductance spectroscopy measurements of $G_2$ and $G_3$ over a wide range of $V_\mathrm{W23}$. The voltages on the tunnel barrier gates $V_\mathrm{P2}$ and $V_\mathrm{P3}$ were compensated according to Eqs. \ref{eq:long_plunger}. The variable on the vertical axis is therefore marked $\tilde V_\mathrm{W23}$ instead of $V_\mathrm{W23}$. No subgap state crosses zero bias for gate voltages $\tilde V_\mathrm{W23}<\SI{-3.3}{\volt}$. A line cut at zero bias for $G_2$ and $G_3$ is plotted in (c). A zoomed-in measurement of $G_2$ and $G_3$ is depicted in (d) and (e). It reveals subgap states whose peak positions show identical nontrivial gate voltage dependence in both $G_2$ and $G_3$. The data for $G_2$ and $G_3$ at zero bias is plotted in (f) for comparison.}
\end{figure}

The data shown in the following was measured on device 1 and provides additional information to the discussion in the main text. Figures~\ref{fig:evolution}(a,b) show the evolution of $G_2$ and $G_3$ from $V_\mathrm{W23}=\SI{-3.8}{\volt}$ to $V_\mathrm{W23}=\SI{-3.1}{\volt}$ at $B_{||}=\SI{1.6}{\tesla}$. Note that while changing $V_\mathrm{W23}$, the voltages $V_\mathrm{P2}$ and $V_\mathrm{P3}$ were changed according to the relations

\begin{equation}
 \begin{aligned} \label{eq:long_plunger}
     V_\mathrm{P2}&=-\SI{10}{\milli\volt}-\frac{0.05}{0.40} \cdot(V_\mathrm{W23}+\SI{3.1}{\volt})\\
     V_\mathrm{P3}&=-\SI{135}{\milli\volt}-\frac{0.045}{0.40} \cdot (V_\mathrm{W23}+\SI{3.5}{\volt}) 
 \end{aligned}
\end{equation}
The lines given by these relations are shown in green in Figs.2(b, c) of the main text. This compensates for the effect of the gate $V_\mathrm{W23}$ on the tunnel barriers and ensures that the high bias conductance stays around $\sim 0.1\; e^2/h$ throughout the complete range of $\SI{-3.8}{\volt}<V_\mathrm{W23}<\SI{-3.1}{\volt}$. To denote that more than one gate voltage was changed during the measurement, we label the variable $\tilde V_\mathrm{W23}$ instead of $V_\mathrm{W23}$.
There are no states crossing zero bias for voltage values $\tilde V_\mathrm{W23}<\SI{-3.3}{\volt}$. Above that value Andreev bound states (ABSs) cross the gap. Line cuts at zero bias are displayed in Fig.~\ref{fig:evolution}(c). A measurement with higher resolution within the range $\SI{-3.06}{\volt}<\tilde V_\mathrm{W23}<\SI{-3.12}{\volt}$ in Figs.~\ref{fig:evolution}(d, e) shows that subgap states evolve with the same nontrivial dependence on $\tilde V_\mathrm{W23}$ in measurements of $G_2$ and $G_3$. At the same time, the subgap states can appear with different strength in conductance in $G_2$ and $G_3$.  A line cut taken at zero bias in Fig.~\ref{fig:evolution}(f) shows the correlated $\tilde V_\mathrm{W23}$ dependence on both sides in the form of coinciding peak positions, while the difference in conductance value by up to one order of magnitude for some of the subgap states is apparent from the difference in peak height.

\subsection{\label{sec:peakextraction}Extraction of peak positions}
\begin{figure}
\includegraphics[scale=0.9]{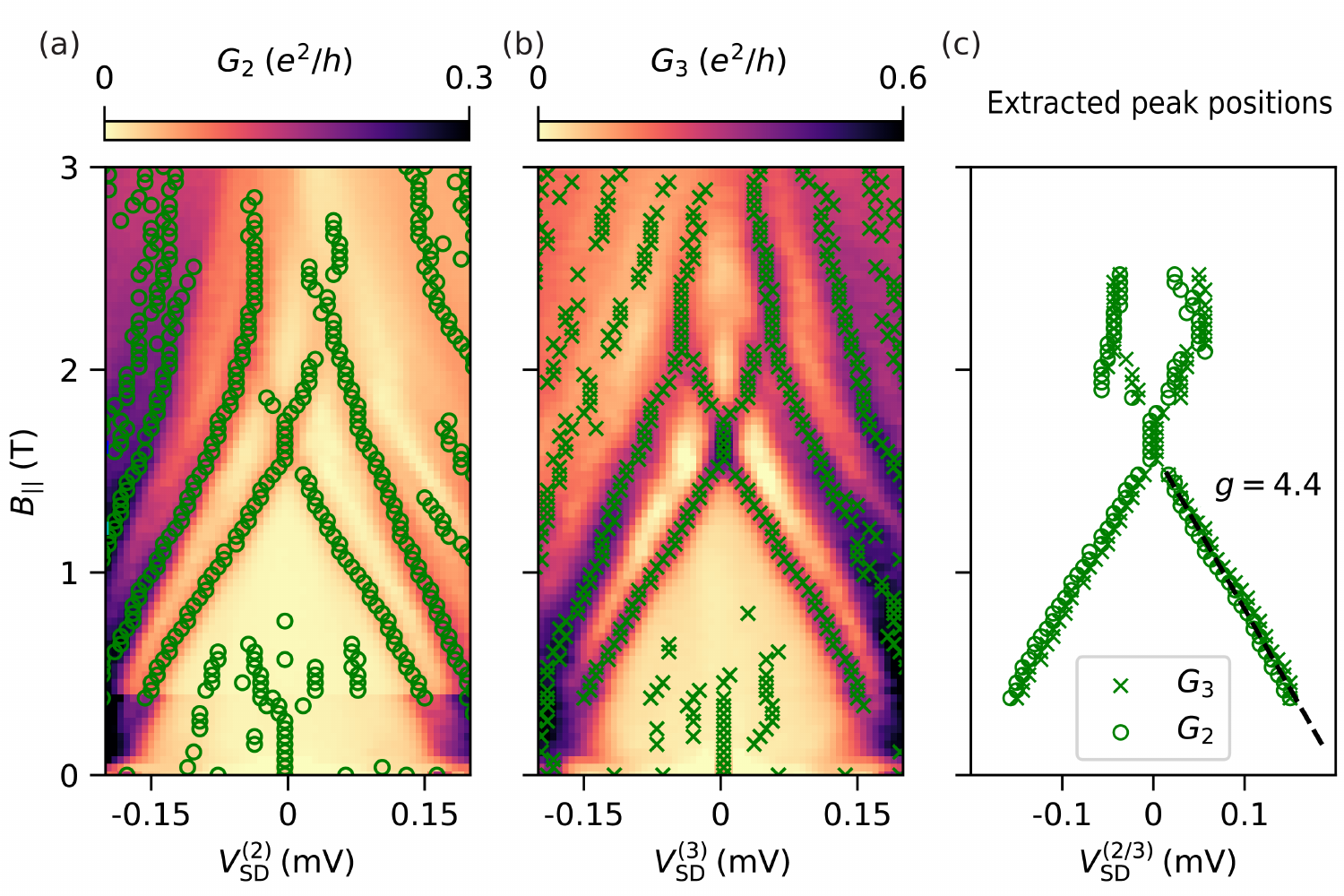}
\caption{\label{fig:peak_extraction}(a, b) Identical data to the ones in Figs.~3(a, b) in the main text. The green markers indicate the position of identified local maxima for individual magnetic field values. They originate from subgap states, noise, and a supercurrent peak at low magnetic field values. (c) Locations of the local maxima from (a) and (b) after removing the peaks that we identify not to stem from the low energy subgap states of interest.}
\end{figure}

We extracted peak positions of subgap states from  $G_2$ and $G_3$ as functions of magnetic field and $V_\mathrm{SD}^{(2)}$ and $V_\mathrm{SD}^{(3)}$ used in Figs.~3(b, c) in the main text as follows: local maxima of $G_{(2/3)}$ were found for each value of $B_{||}$ within the range $\SI{0.38}{\tesla}<B_{||}<\SI{2.5}{\tesla}$, wherever the subgap state is resolved. Detected peaks are marked in the measurements of $G_2$ and $G_3$ in Figs.~\ref{fig:peak_extraction}(a, b). As a next step, the maxima that are obviously (by eye) not arising from the subgap state of interest but due to noise or higher excited states are discarded. The same applies for values of $B_{||}$ at which only one subgap peak could be detected in one of the two measurements, while other measurement showed two peaks. These post-selected peak positions associated with the subgap state of interest are plotted in Fig.~\ref{fig:peak_extraction}(c).\\

\subsection{\label{sec:control_experiment}No nonlocal signature in the absence of Andreev bound states}
In the case of a rather negative gate voltage $V_\mathrm{W23}$ around \SI{-4}{\volt} we observed no subgap state that appears in both $G_2$ and $G_3$, while the quantum dot (QD) resonance in the tunnel barrier under $\mathrm{P_2}$ was still present. In this configuration, no nonlocal signatures can be measured in tunneling spectroscopy.

\begin{figure}
\includegraphics[scale=0.9]{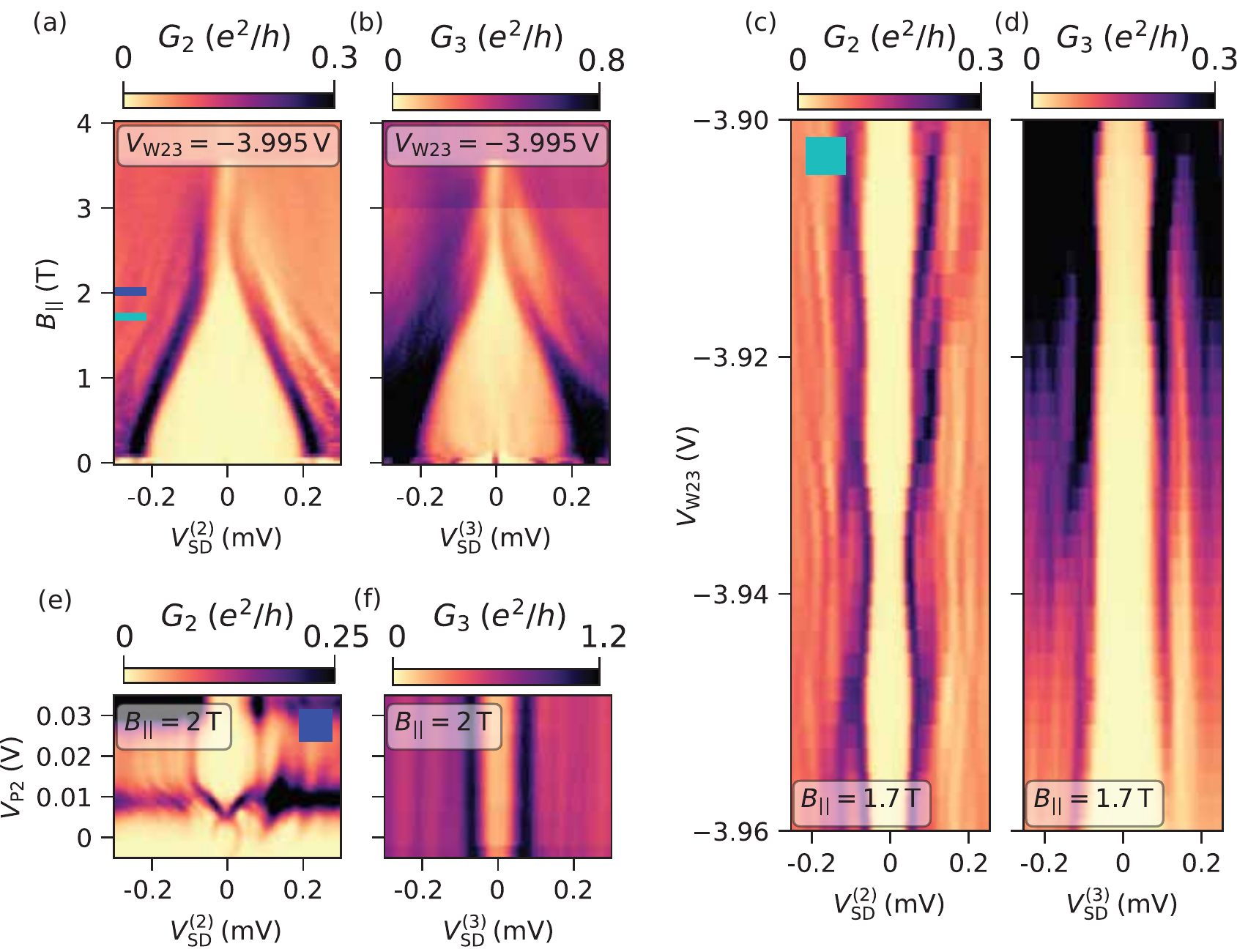}
\caption{\label{fig:control} (a) and (b) show tunneling spectroscopy data of $G_2$ and $G_3$ with $V_\mathrm{W12}=V_\mathrm{W34}=\SI{-7}{\volt}$ and $V_\mathrm{W23}=\SI{-3.995}{\volt}$. There is no subgap state visible. Tunneling spectroscopy with respect to changing gate voltage $V_\mathrm{W23}$ at $B_{||}=\SI{1.7}{\tesla}$ is plotted in (c) and (d). There is no subgap state present. Tunneling conductance measurement with respect to the gate voltage $V_\mathrm{P2}$ that tunes the tunnel barrier is depicted in (e) and (f). While $G_2$ shows a clear subgap state at the position of the QD resonance, the data of $G_3$ in (f) shows no measurable change at the same gate voltage.}
\end{figure}

Tunneling spectroscopy in Figs.~\ref{fig:control}(a, b) shows no subgap states crossing zero bias. A measurement of tunneling spectroscopy in a small range around $V_\mathrm{W23}=\SI{-3.93}{\volt}$ at $B_{||}=\SI{1.6}{\tesla}$ in Figs.~\ref{fig:control}(c, d) shows a gap with no subgap states persisting over the whole range of  $V_\mathrm{W23}$. In Fig.~\ref{fig:control}(e) tunneling spectroscopy as a function of the gate voltage $V_\mathrm{P2}$ reveals the QD resonance around $V_\mathrm{P2}=\SI{0.01}{\volt}$ at high bias \cite{chang_YSR, kasper_tuningYSR, lee_singletdoublet, alex_QDinLoop}.
The measurement of $G_3$ while sweeping $V_\mathrm{P2}$ over the QD resonance is shown in Fig.~\ref{fig:control}. Throughout the measurement, $G_3$ is independent of $V_\mathrm{P2}$, in particular around the value $V_\mathrm{P2}=\SI{0.01}{\volt}$ at which the QD resonance appears in $G_2$.

In summary, in the absence of a discrete subgap state extending between the two probes, no change in the density of states is measured on one probe while a QD on the other tunnel probe is brought on resonance. This shows that only an extended ABS can hybridize with a local QD and lead to a measureable nonlocal signature on a tunnel probe $\SI{0.6}{\micro\meter}$ away from the QD.

\section{\label{sec:S2}Supplementary data on device 1 - Andreev bound states under $\mathbf{W_{12}}$}
In the main text, data of the hybridization between an ABS in the NW segment under the gate $\mathrm{W}_{23}$ and a local QD is shown. Here data of the comparable experiment with an ABS in the NW segment under the gate $\mathrm{W}_{12}$, which is marked green in Fig.~\ref{fig:P_scan_dev2}(a), is presented. 

\begin{figure}
\includegraphics[scale=0.9]{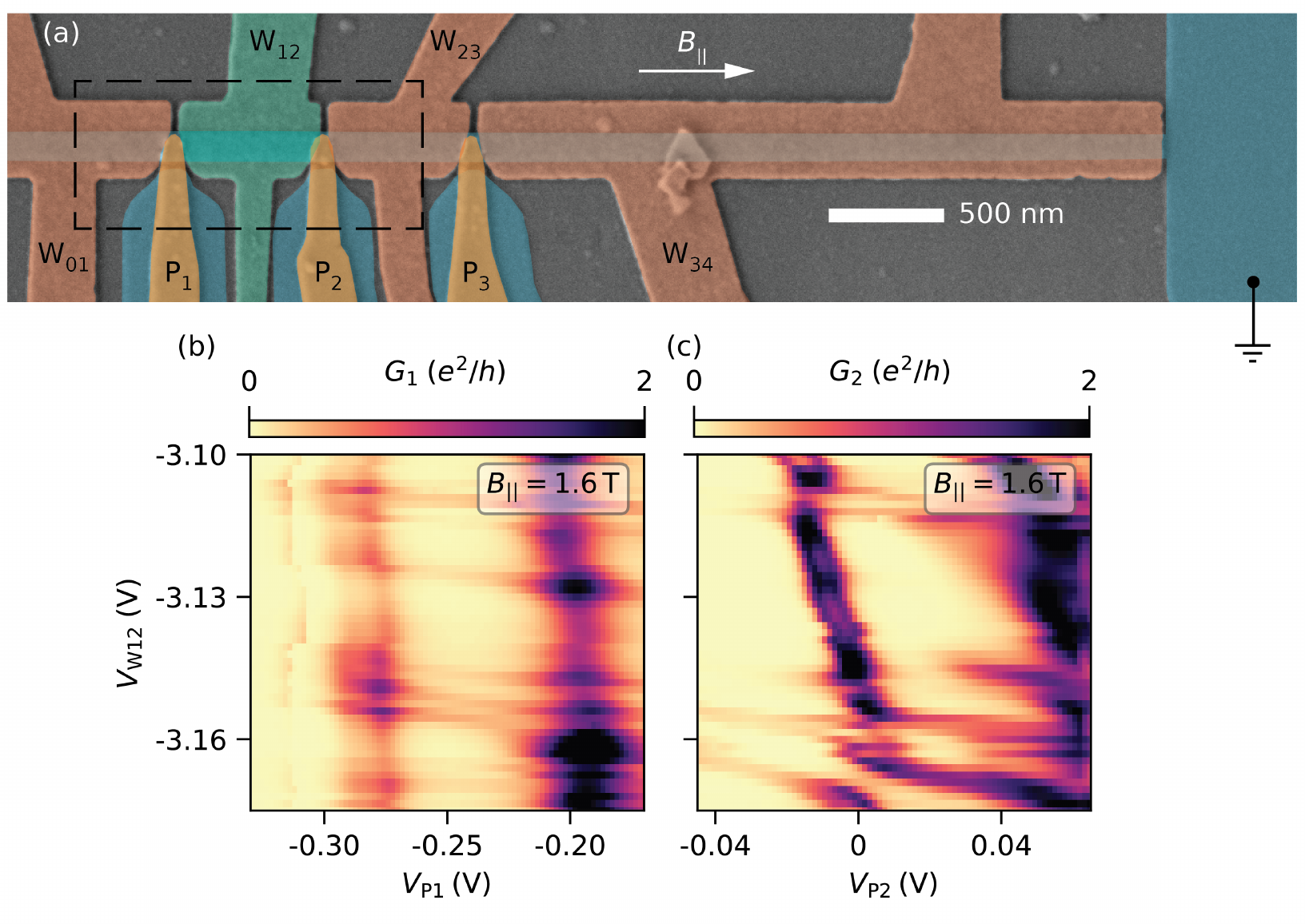}
\caption{\label{fig:sem_left}\label{fig:gatemap_left}(a) False-color scanning electron micrograph of device 1. (b) and (c) show conductance resonances measured at $B_{||}=\SI{1.6}{\tesla}$ at $V^{(j)}_{\mathrm{SD}}=\SI{0}{\volt}$. Both measurements show two distinct families of states - horizontal ones, that strongly depend on $V_\mathrm{W23}$, and vertical ones, that strongly depend on the gate voltage that tunes the respective tunnel barrier.}
\end{figure}

The gate voltage dependence of conductance resonances with respect to the voltages on the gate $V_\mathrm{W12}$ that covers the NW and the two gates $\mathrm{P}_{1}$ and $\mathrm{P}_{2}$, while keeping the neighboring gate voltages at $V_\mathrm{W01}=V_\mathrm{W23}=\SI{-4.5}{\volt}$ can be seen in Fig.~\ref{fig:gatemap_left}(c,d). There are horizontal resonances in $G_2$ and $G_3$ that are associated with extended states in the NW. In addition vertical resonances, which can be attributed to localized states in the respective tunnel barrier regions, are present. 

\begin{figure}
\includegraphics[scale=0.9]{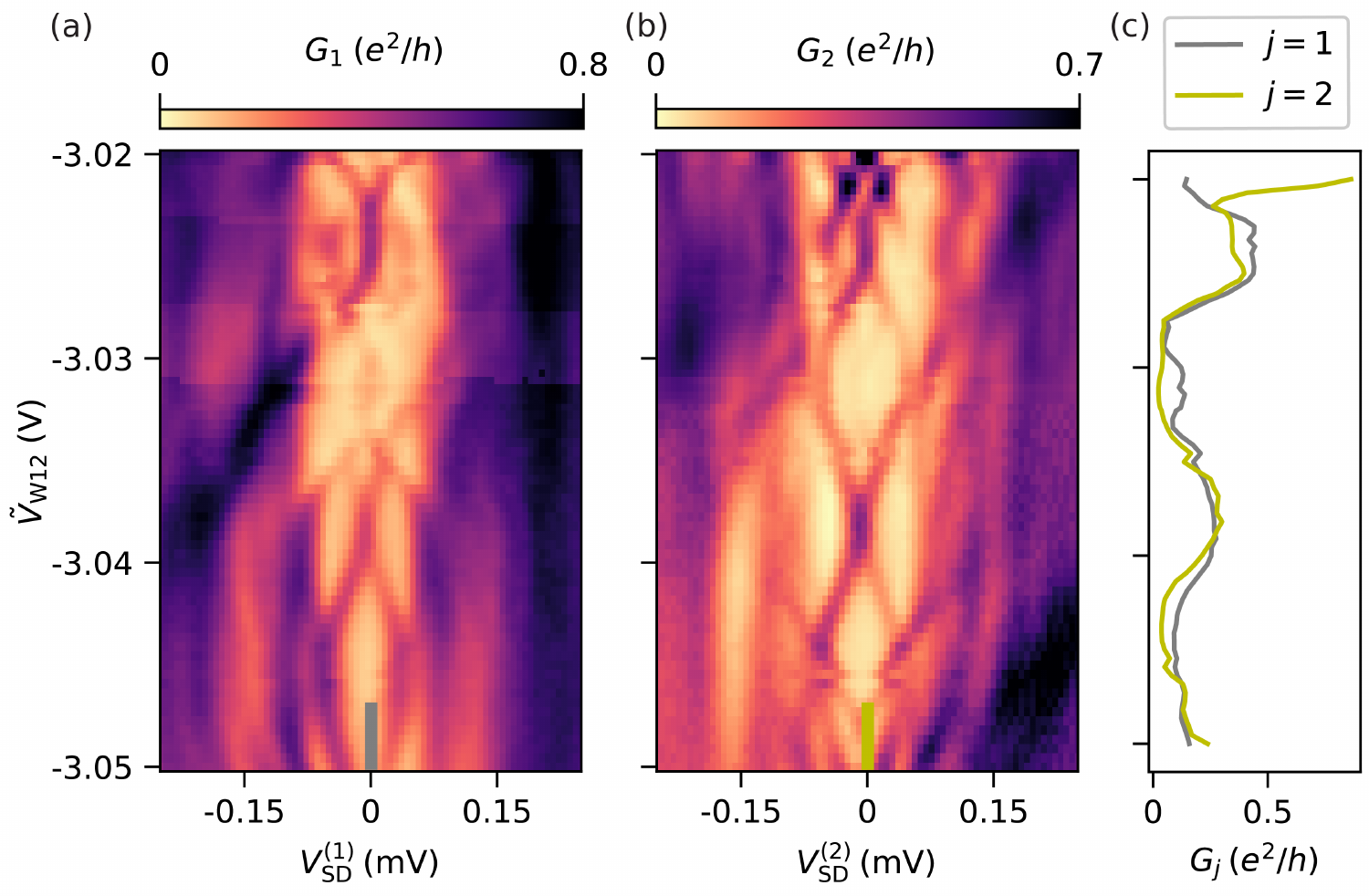}
\caption{\label{fig:plunger_scan_left}(a, b) Tunneling spectroscopy measurements of $G_2$, $G_3$ with respect to $\tilde V_\mathrm{W12}$. The voltages on the other gates were fixed at $V_\mathrm{W01}=V_\mathrm{W23}=V_\mathrm{W34}=\SI{-4.5}{\volt}$. Subgap states appear in both measurements of $G_1$ and $G_2$. The data for $G_2$ and $G_3$ at $V^{(1)}_{\mathrm{SD}}=V^{(2)}_{\mathrm{SD}}=\SI{0}{\volt}$ is plotted in (c) for comparison.}
\end{figure}

Tunneling spectroscopy data as a function of $V_\mathrm{W12}$, using tunnel probes 1 and 2, are shown in Figs.~\ref{fig:plunger_scan_left}(a,b). Voltages on the gates $\mathrm{P}_1$ and $\mathrm{P}_2$ were compensated according to:
\begin{equation}
    \begin{aligned}
     V_\mathrm{P1}&=-\SI{307}{\milli\volt}-\frac{0.01}{0.80} \cdot(V_\mathrm{W12}+\SI{3}{\volt})\\
     V_\mathrm{P2}&=-\SI{15}{\milli\volt}-\frac{0.015}{0.80} \cdot(V_\mathrm{W12}+\SI{3}{\volt}).
    \end{aligned}
\end{equation}
To denote that more than one gate voltage was changed during the measurement, the variable is labeled $\tilde V_\mathrm{W12}$ instead of $V_\mathrm{W12}$.
Tunneling conductances $G_1$ and $G_2$ in Figs.~\ref{fig:plunger_scan_left}(a, b) show subgap states as a function of $\tilde V_\mathrm{W12}$. These states show similar dependences on $\tilde V_\mathrm{W12}$ in the two measurements. This is emphasized by the line cut taken at zero bias in Fig.~\ref{fig:plunger_scan_left}(c).

\begin{figure}
\includegraphics[scale=0.9]{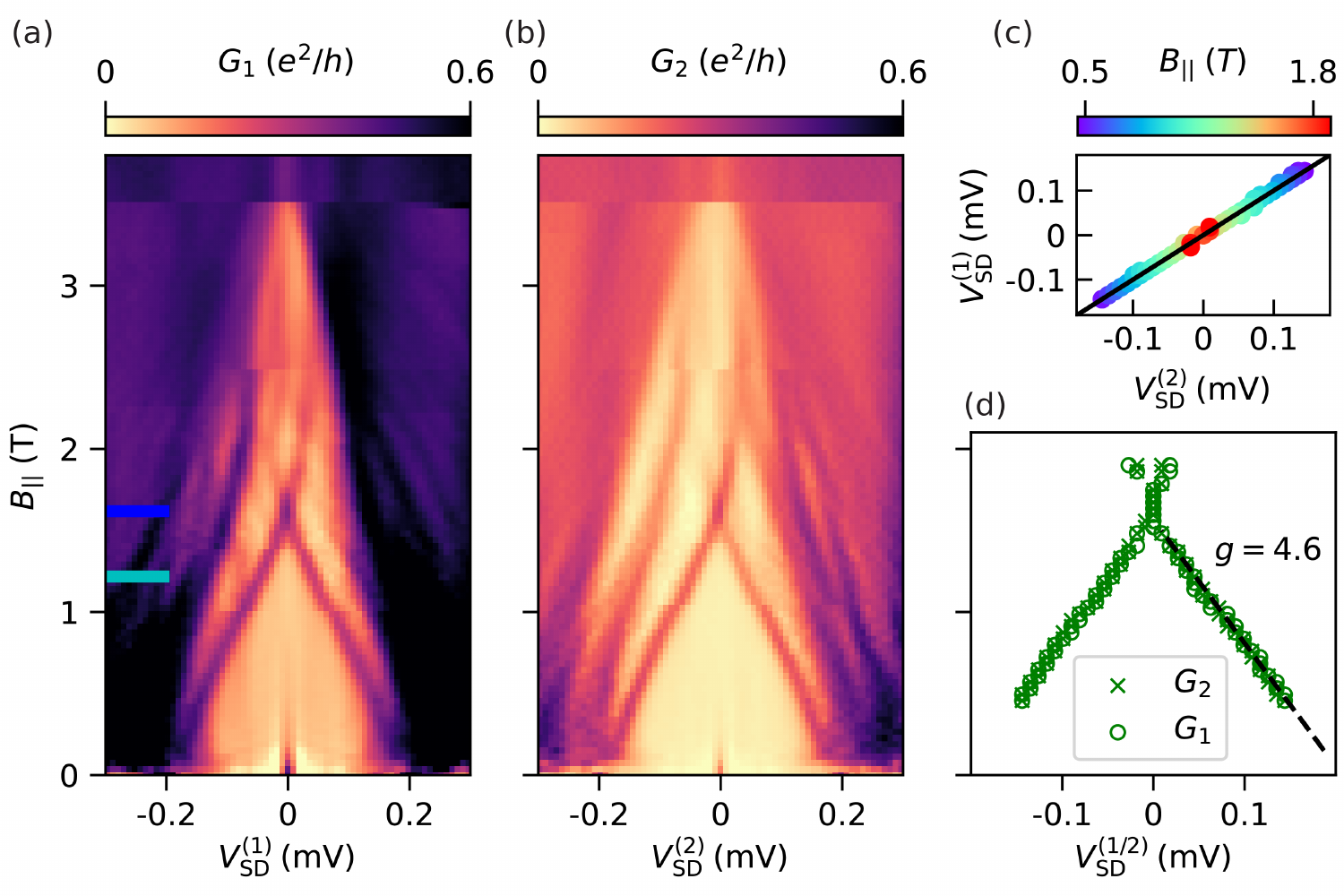}
\caption{\label{fig:field_scan_left} (a, b) Tunneling spectroscopy measurements $G_1$, $G_2$ with respect to magnetic field $B_{||}$ at $V_\mathrm{W12}=\SI{-3.02}{\volt}$. The peak positions that were extracted according to the procedure described in section \ref{sec:peakextraction} are plotted in (d). A parametric plot of the same data points as in (d) is shown in (c).}
\end{figure}

Tunneling conductance as a function of magnetic field $B_{||}$ is shown in Fig.~\ref{fig:field_scan_left}(a, b). Subgap states were found to emerge at low magnetic field from the quasiparticle continuum. At $B_{||}=\SI{1.6}{\tesla}$, subgap states cross zero bias and overshoot at $B_{||}=\SI{2}{\tesla}$. For every value of $B_{||}$ we extracted the peak position from the measured trace of $G_1(G_2)$ according to the procedure outlined in \ref{sec:peakextraction}. Extracted peak positions are shown in Fig.~\ref{fig:fieldscan_dev2}(d). The same peak positions are plotted parametrically in Fig.~\ref{fig:fieldscan_dev2}(c). In the parametric plot, all points lie on or close to the identity line demonstrating the similar behavior of the subgap states in measurements of $G_1$ and $G_2$. Due to their correlated behavior in $G_1$ and $G_2$ with respect to magnetic field and $V_\mathrm{W12}$ gate voltage, we attribute these subgap states to ABSs that extend over the NW segment under gate $V_\mathrm{W12}$.\\

\begin{figure}
\includegraphics[scale=0.9]{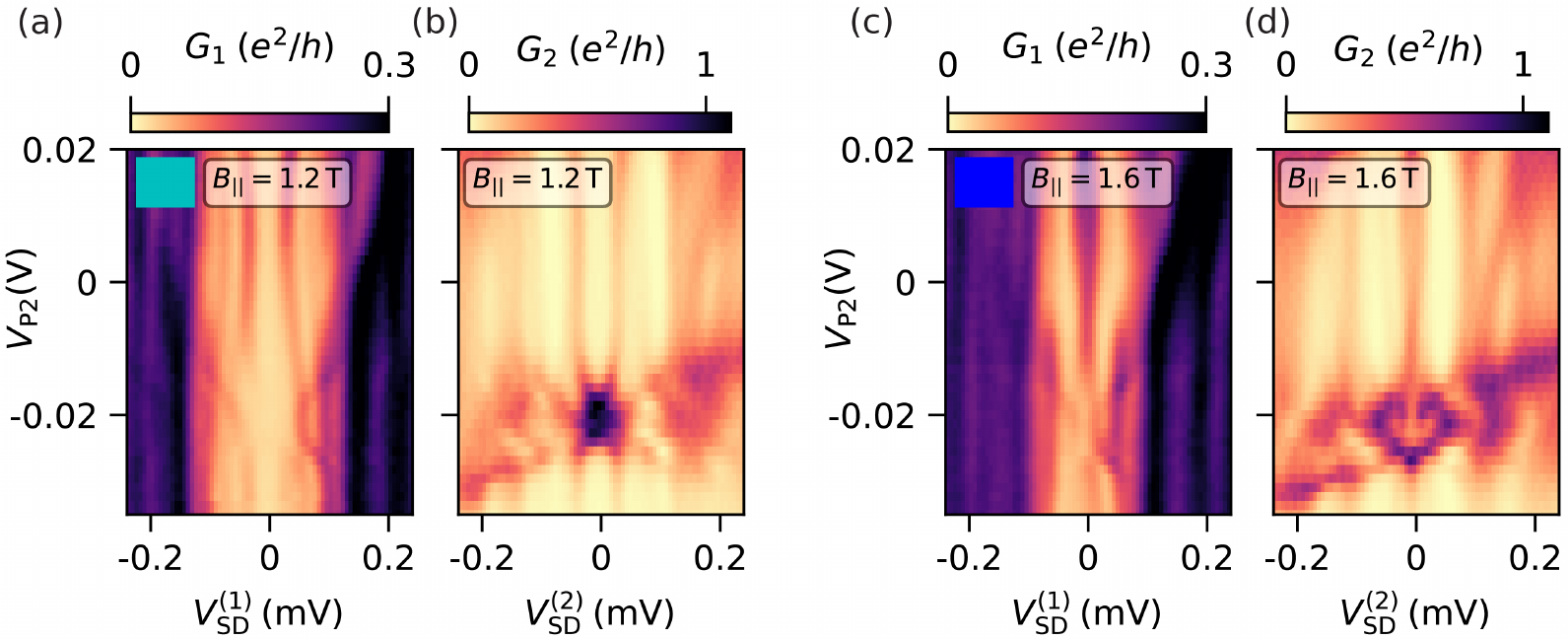}
\caption{\label{fig:pc_left}Tunneling spectroscopy data at $V_\mathrm{W12}=\SI{-3.02}{\volt}$ as a function of gate voltage $V_\mathrm{P2}$ is plotted in (a) and (b). A QD resonance is clearly visible in the measurement of $G_2$. The same measurement at higher magnetic field $B_{||}=\SI{1.6}{\tesla}$ is shown in (c) and (d).}
\end{figure}

To investigate the effect on the subgap states as the ABS is brought on resonance with the QD in the tunnel barrier under $\mathrm{P}_2$, we performed tunneling spectroscopy with respect to the voltage on gate $\mathrm{P}_2$. The measurement of $G_1$ and $G_2$ at $B_{||}=\SI{1.1}{\tesla}$ is plotted in Figs.~\ref{fig:pc_left}(a) and (b). A QD resonance can be seen around $V_\mathrm{P2}=\SI{-0.02}{\volt}$ as signal enhancement at high bias in the measurement of $G_2$. In the vicinity of the QD resonance, the subgap states exhibit a `bowtie' shape with a conductance enhancement at the point where the states cross zero bias in $G_2$. The simultaneous measurement of tunneling conductance $G_1$ shows that the subgap states are almost constant in energy, up to a small region around $V_\mathrm{P2}\sim\SI{-0.02}{\volt}$ around the QD resonance. In the vicinity of the QD resonance, the lowest energy subgap state decreases in intensity in $G_1$ such that it is not visible at the position of the QD resonance.

A measurement of the same type at a higher magnetic field value of $B_{||}=\SI{1.6}{\tesla}$ is shown in Figs.~\ref{fig:pc_left}(c) and (d). The quantity $G_2$, which is measured through the QD, shows the QD resonance crossing the gap at $V_\mathrm{P2}=\SI{-0.02}{\volt}$. A similar pattern to the one shown in Fig. 4(c) in the main text is visible around zero bias at the point where the QD is on resonance with the subgap state close to zero bias. The resolution is not sufficient to resolve the exact pattern and determine whether the subgap states shift in energy around the QD resonance to form a `diamond' shape. The measurement of $G_1$ shows that the conductance value measured for the subgap state at zero bias decreases abruptly around the QD resonance while a splitting of the states away from zero energy is resolved. 

For all measurements in Fig.~\ref{fig:pc_left} a clear effect on the ABSs can be observed at tunnel barrier 1 when the subgap state is brought on resonance with a QD in tunnel barrier 2, even though the precise line shape of the subgap state is not resolved in this measurement. The change in conductance value of the subgap state at the point of the QD at both ends is stronger in this measurement compared to the data presented in the main text.

\section{\label{sec:S3}Device 2}

\begin{figure}
\includegraphics[scale=1.0]{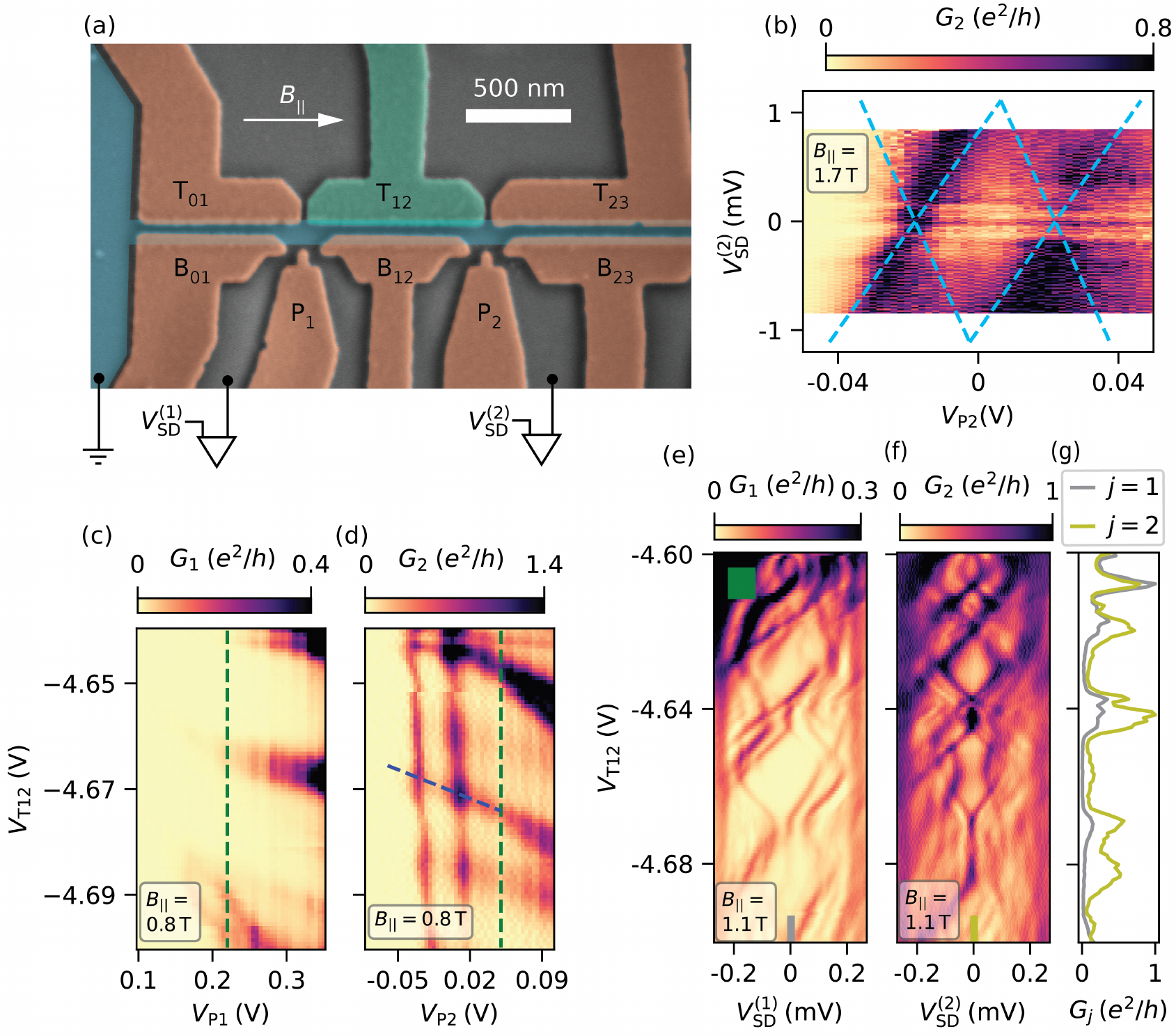}
\caption{\label{fig:gate_coupling_dev2}\label{fig:sem_dev2}\label{fig:P_scan_dev2}(a) False-color scanning electron micrograph of device 2. The device consists of a superconducting Al NW on top of a InAs quantum well. The gates labeled $V_{\mathrm{P}j}$ can be used to tune the tunnel barrier between the tunnel probe $j$ and the wire ($j \in \{1,2,3\}$). The gates labled $V_{\mathrm{B}kl}$ form the confining potential of the quasi one-dimensional NW at the bottom edge while also forming tunnel barriers. The gates labeled  $\mathrm{T}_{kl}$ confine the NW at the top edge and tune the chemical potential of individual NW segments ($kl \in \{01,12,23\}$). (c, d) Tunneling conductance measurements of $G_1$, $G_2$ with $V^{(j)}_{\mathrm{SD}}=\SI{0}{\volt}$. States that couple strongly to the gate voltage $V_\mathrm{T12}$ are visible in both measurements. Two sharp resonances that strongly depend on the gate voltage $V_\mathrm{P2}$ appear in the measurement of $G_2$. (b) Tunneling spectroscopy at $V_\mathrm{T12}=\SI{-4.67}{\volt}$. The dashed lines serve as a guide to the eye. Tunneling spectroscopy at $B_{||}=\SI{1.1}{\tesla}$, when moving the gate voltages along the green dashed lines at $V_\mathrm{P1}=\SI{0.22}{\volt}$  and at $V_\mathrm{P2}=\SI{0.05}{\volt}$ in (c, d) are plotted in (e, f). While the absolute values of tunneling conductance in $G_1$ and $G_2$ are different, the structure of the states in energy is nearly identical. A line cut at $V^{(1)}_{\mathrm{SD}}=V^{(2)}_{\mathrm{SD}}=\SI{0}{\volt}$ of the data in (e, f) is shown in (g).}
\end{figure}

We fabricated a second device with semiconducting, instead of superconducting, side probes, i.e, with the epitaxial Al removed in the leads of the side probes. The device is depicted in Fig.~\ref{fig:sem_dev2}(a). It consists of a superconducting strip of Al on top of an InAs quantum well. The gates labeled $\mathrm{T}_{kl}$ are used to electrostatically define the NW on one side and tune the electron density of individual segments of the NW. The gates labeled $\mathrm{B}_{kl}$ serve the purpose of defining the wire on the other side, tuning its electron density, and forming a quantum point contact between adjacent $\mathrm{B}_{kl}$ gates. The gates $\mathrm{P}_1$ and $\mathrm{P}_2$ add additional control over the quantum point contacts. In the data presented in the following, the gate voltages $V_\mathrm{T01}=V_\mathrm{T23}=\SI{-6}{\volt}$, $V_\mathrm{B01}=V_\mathrm{B23}=\SI{-2.4}{\volt}$, $V_\mathrm{B12}=\SI{-2.5}{\volt}$ were kept at fixed values.\\ 

Measurements of conductance resonances in $G_1$ ($G_2$) at zero bias and magnetic field $B=\SI{0.8}{\tesla}$ with respect to the gate voltages $V_\mathrm{T12}$ and $V_\mathrm{P1}$ ($V_\mathrm{P2}$) are shown in Figs.~\ref{fig:gate_coupling_dev2}(c) and (d). Both measurements reveal resonances that strongly depend on $V_\mathrm{T12}$ originating from states inside the NW. The measurement of $G_2$ furthermore shows two resonances that strongly couple to the gate voltage $V_\mathrm{P2}$. These features can be associated with a QD in the tunnel barrier, which is verified by bias spectroscopy shown in Fig.~\ref{fig:gate_coupling_dev2}(b). The conductance shows a Coulomb diamond at high bias as emphasized by the cyan, dashed lines. This allows for an estimate of the charging energy to be $\sim \SI{1}{\milli\eV}$. 

Tunneling spectroscopy measurements of subgap states in $G_1$ and $G_2$ at $B_{||}=\SI{1.1}{\tesla}$ are shown in Fig.~\ref{fig:P_scan_dev2}(e) and (f). The gate voltages $V_\mathrm{P1}=\SI{0.22}{\volt}$ and $V_\mathrm{P2}=\SI{0.05}{\volt}$ were kept constant during this measurement, which corresponds to the green dashed lines in Fig.~\ref{fig:P_scan_dev2}(c, d). Similar to the results of device 1, states appear in both measurements of $G_1$ and $G_2$ with the same nontrivial dependence on the voltage $V_\mathrm{T12}$. The conductance value with which states appear in the two measurements can be very different, however. A line cut at zero bias is depicted in Fig.~\ref{fig:P_scan_dev2}(g) and captures the correlated dependence of conductance peaks with respect to $V_\mathrm{T12}$ in the form of concomitant peak positions. The difference in conductance values results in a difference in peak heights for the two different curves in Fig.~\ref{fig:P_scan_dev2}(g). \\

\begin{figure}
\includegraphics[scale=0.9]{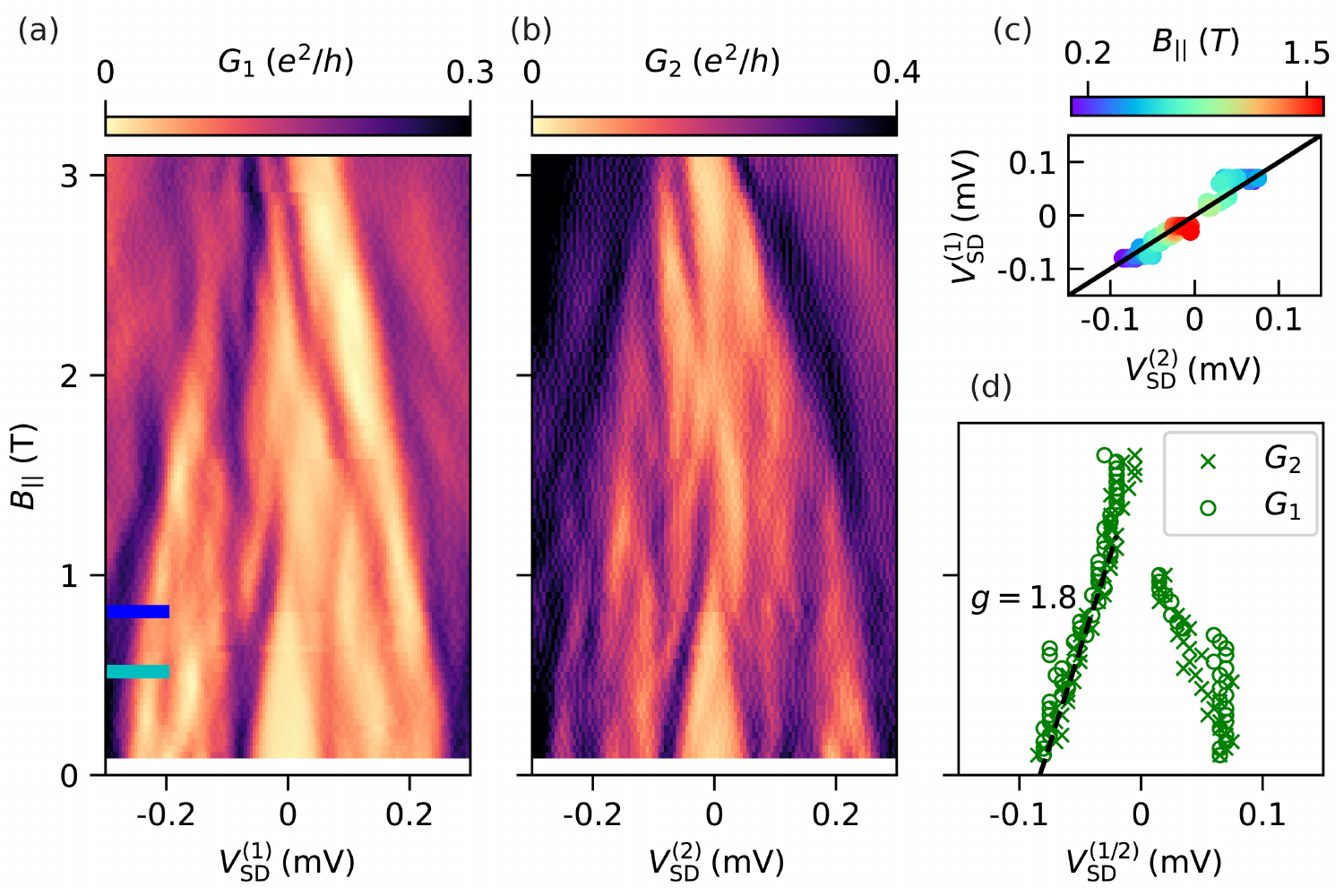}
\caption{\label{fig:fieldscan_dev2} (a,b) Tunneling spectroscopy with respect to magnetic field $B_{||}$ with $V_\mathrm{T12}=\SI{-4.67}{\volt}$. The induced gap at zero field is $\sim \SI{80}{\micro\eV}$. Peak positions extracted from $G_1$ and $G_2$ associated with the lowest energy state are plotted in (d). (c) Parametric plot of the peak positions from (d).}
\end{figure}

Figures~\ref{fig:fieldscan_dev2}(a, b) show tunneling spectroscopy data of $G_1$ and $G_2$ for device 2 as a function of magnetic field $B_{||}$ parallel to the NW. The induced gap at low field given by the lowest energy states is $\sim \SI{80}{\micro\eV}$. Note that this device was fabricated on material where the InAs quantum well is separated by an InAlAs barrier from the superconducting Al. 

The lowest energy states in both $G_1$ and $G_2$ cross zero bias at around $B_{||}=\SI{1.5}{\tesla}$. Extracting the peak positions for each value of $B_{||}$ and plotting them parametrically reveals strong correlation as shown in Fig.~\ref{fig:fieldscan_dev2}(c). Based on this strong correlation in $G_1$ and $G_2$ with respect to magnetic field and $V_\mathrm{T12}$, we attribute these subgap states to extended ABSs in the NW.\\

\begin{figure}
\includegraphics[scale=1.0]{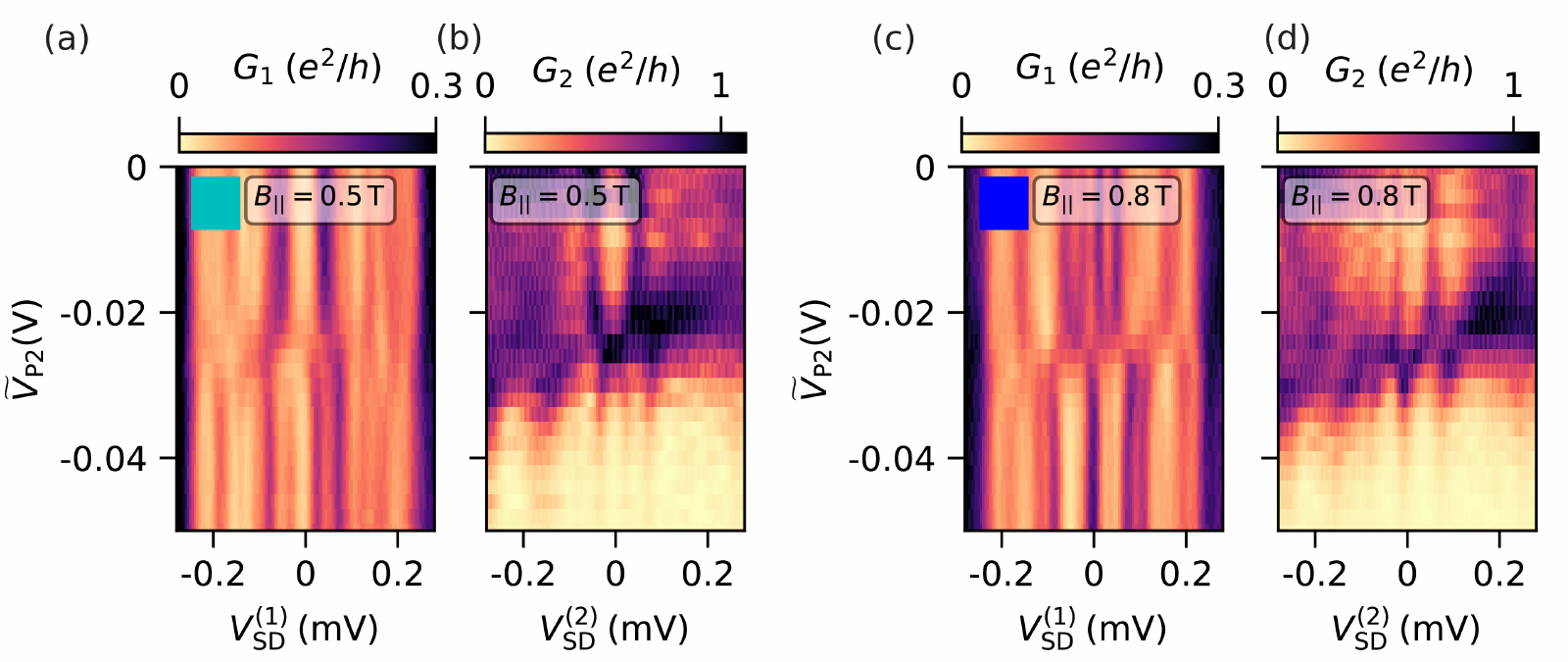}
\caption{\label{fig:pc_dev2}(a, b) Tunneling conductance measured while changing the gate voltages $V_\mathrm{P2}$ and $V_\mathrm{T12}$ along the blue dashed line according to equation \ref{eq:pradaclarke_compensation}. To indicate that more than one gate voltage was changed, the variable on the vertical axes is labeled $\tilde V_\mathrm{P2}$ instead of $V_\mathrm{P2}$. At $\tilde V_\mathrm{P2} = \SI{-0.025}{\volt}$ a QD resonance is visible in $G_2$. In both $G_1$ and $G_2$, the subgap states draw a `bowtie' shape at this value of gate voltage. (c, d) The same measurement as in (a, b) but at a higher magnetic field value of $B_{||}=\SI{0.8}{\tesla}$. In $G_1$ one can clearly see two peaks merging into a zero-bias conductance peak at the position of the QD resonance. }
\end{figure}

To investigate the behavior around places where the ABSs in the NW are on resonance with the QD in tunnel barrier 2, we measured tunneling spectroscopy with respect to $V_\mathrm{P2}$. While sweeping $V_\mathrm{P2}$, the gate voltage $V_\mathrm{T12}$ was compensated to mitigate the effect that $V_\mathrm{P2}$ has on the ABSs inside the NW. The parametric equation that was used is given by 
\begin{align}
    \label{eq:pradaclarke_compensation}
    V_\mathrm{T12}&=-\SI{4.664}{\volt}-\frac{0.014}{0.18}\cdot (V_\mathrm{P2}+\SI{0.08}{\volt})
\end{align}
and is depicted by the blue dashed line in Fig.~\ref{fig:gate_coupling_dev2}(d). To denote that more than one gate voltage was changed during the measurement, the variable is labeled $\tilde V_\mathrm{P2}$ instead of $V_\mathrm{P2}$. 
The measured tunneling spectroscopy $G_1$ and $G_2$ is depicted in Fig.~\ref{fig:pc_dev2}(a) and (b) for a magnetic field value of $B_{||}=\SI{0.5}{\tesla}$. The lowest energy subgap states in $G_2$ show a clear crossing at zero energy with the characteristic `bowtie' shape at the location of the QD resonance at $\tilde V_\mathrm{P2}=\SI{-0.02}{\volt}$. The states at higher bias are not well resolved in the measurement of $G_2$. The lowest energy subgap states in $G_1$ show a `bowtie' shape in the form of a zero bias crossing at the location of the QD resonance. For the measurement of $G_1$, the higher excited states are well resolved and a shift in energy for all higher excited states is visible at the position of the QD resonance. 

Identical measurements of $G_1$ and $G_2$ at a higher magnetic field, $B_{||}=\SI{0.8}{\tesla}$, are depicted in Figs.~\ref{fig:pc_dev2}(c, d). At the point of the QD resonance all conductance resonances shift in energy. The pair of subgap states with lowest energy merge from two separated peaks at $\tilde V_\mathrm{P2}>\SI{-0.02}{\volt}$ to a single peak at zero bias for $\tilde V_\mathrm{P2}<\SI{-0.02}{\volt}$. This pattern shows close similarity to the results presented in Ref.~\cite{mingtang_science}.

\end{document}